# A Novel Design of IEEE 802.15.4 and Solar Based Autonomous Water Quality Monitoring Prototype using ECHERP

Fredrick Romanus Ishengoma

*College of Informatics and Virtual Education, The University of Dodoma, Dodoma, Tanzania.*

*ishengomaf@gmail.com*

**Abstract**

The recently advancement in Wireless Sensor Network (WSN) technology has brought new distributed sensing applications such as water quality monitoring. With sensing capabilities and using parameters like pH, conductivity and temperature, the quality of water can be known. This paper proposes a novel design based on IEEE 802.15.4 (Zig-Bee protocol) and solar energy called Autonomous Water Quality Monitoring Prototype (AWQMP). The prototype is designed to use ECHERP routing protocol and Adruino Mega 2560, an open-source electronic prototyping platform for data acquisition. AWQMP is expected to give real time data acquirement and to reduce the cost of manual water quality monitoring due to its autonomous characteristic. Moreover, the proposed prototype will help to study the behavior of aquatic animals in deployed water bodies.

**Keywords: ECHERP, IEEE 802.15.4 technology, Solar and Water Quality Monitoring.**

## I. Introduction

About 2.6 billion people – half the developing world – lack even a simple 'improved' latrine and 1.1 billion people have no access to any type of improved drinking source of water (WHO, 2013). The effect of this is millions people dying annually due to diseases caused by lack of drinking safe water and sanitation.

Water quality can be defined as a measure of the suitability of water for a particular use based on selected physical, chemical, and biological characteristics (Bartram, 1996).

The advancement of wireless sensors network (WSN) technology presents a greater opportunity for developing systems that will help in monitoring water quality and sanitation. In order to determine the quality of water, features of water are first examined. The features found in water are then matched to numeric standards to know if the quality of water is appropriate or inappropriate for a particular use.

This study proposes a novel Autonomous Water Quality Monitoring Prototype (AWQMP) based on IEEE 802.15.4 and solar using Equalized Cluster Head Election Routing Protocol (ECHERP). AWQMP design comprises of sensor nodes dispersed in water and a base station located in shore/out of water bodies. Sensor nodes gather numerous data from water using parameters like pH value, turbidity, conductivity, dissolved oxygen, chlorophyll-a, temperature etc. that can be used to determine the quality of water. The prototype collects water data in real time, process and transmits data to the base station via a WSN channel for analysis.

The benefits of using the proposed prototype for water quality monitoring includes real time data acquisition, minimizing the cost of manual monitoring system and studying the behavior of aquatic animals in deployed water bodies.

The rest of the paper is organized as follows: We start by providing related work in section 2. Section 3 is dedicated to the background of IEEE 802.15.4 technology. Section 4 presents factors that influence the design of a WSN system. Section 5 proposes our novel design of IEEE 802.15.4 Based Autonomous Water Monitoring Prototype. Hardware design is presented in section 6 and section 7 describes software design. We conclude and present our future work in section 8.

## II.    Related Work

There exist numerous studies that relates to our work. In this section we presents an overview of research works that are closely related to ours.

Haron et al., (2009), proposed a remote water quality monitoring for prawn farming pond. The proposed system is leveraging on wireless sensors in detecting the water quality and Short Message Service (SMS) technology in delivering alert to the farmers upon detection of degradation of the water quality (2009).

Sutar (2013) developed a wireless sensor node for water quality monitoring in intensive fish quality culture. Zhou et al., (2010) studied the design of the water quality monitoring system for inland lakes based on remote sensing data. The study aimed to reflect the trend of inland lakes environment quality by making full use of advantages of remote sensing data, combined with ground-based observation data.

The study by Park et al., (2006) proposed an integrated technique that uses a genetic algorithm (GA) and a Geographic Information System (GIS) for the design of an effective water quality-monitoring network in a large river system.

Another study work aimed at developing a WSN featuring "plug and play" sensor platforms, novel sensors and lower power consumption communication (SmartCoast, 2009).

Zennarro et al., (2009) presented the design of a water quality measuring system and propose a prototype implementation of a water quality wireless sensor network (WQWSN) building upon the SunSPOT technology. The study aimed to automate the monitoring of drinking water quality in Malawi.

Another recent study by Rao et al. (2013), described the design and demonstration of a low-cost, continuous water-quality monitoring system prototype. The system uses low-cost sensors and open- source hardware aimed at providing continuous water quality measurements at lower cost.

Alkandari et al., (2011) studied the wireless sensor network (WSN) for water monitoring system in Kuwait beaches. The study attempted to deploy the sensors on the sea surface to monitor water characteristics such as temperature, pH, dissolved oxygen, etc., and provide various convenient services for end users who can manage the data via a website with spreadsheet from a long distance or applications in a console terminal.

In this study we design a novel IEEE 802.15.4 and solar-based Autonomous Water Quality Monitoring Prototype (AWQMP) using ECHERP. We use ECHERP routing protocol because it is very efficient in terms of lifetime of the WSN compared to other WSN routing protocols.

## III. IEEE 802.15.4/ZigBee

In this section we presents an overview of the IEEE 802.15.4 (ZigBee) technology. IEEE 802.15.4 describes a communication layer at level 2 in the Open System Interconnection (OSI) model that offers network infrastructure specified for low-powered networks. IEEE 802.15.4 defines the physical and MAC layers.

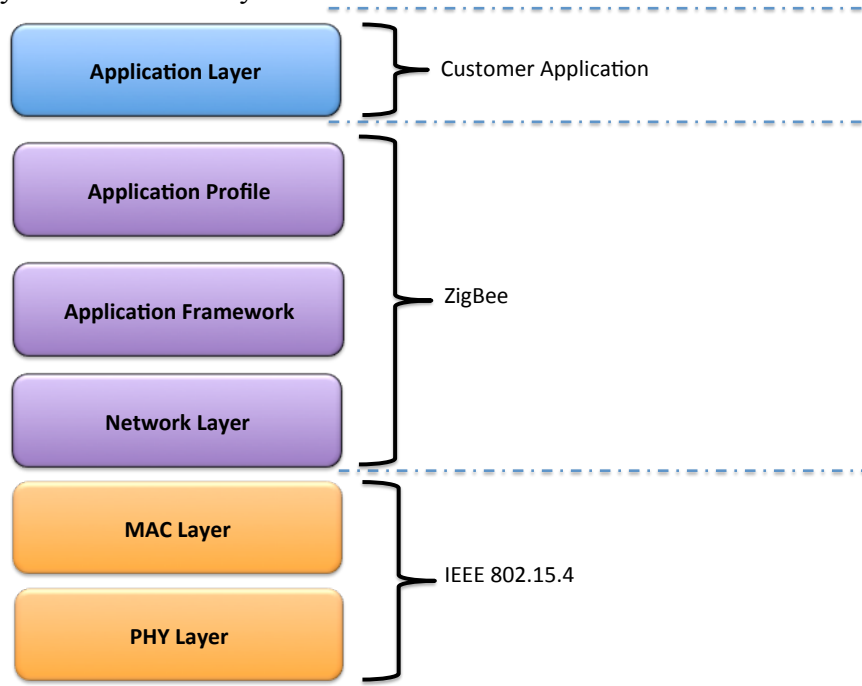

**Figure. 1.** IEEE 802.15.4/ZigBee Architecture

The physical (PHY) layer describes the operation of the Direct Sequence Spread Spectrum (DSSS) radio hardware in 2.4GHz and sub-1GHz band. MAC layer describes access to the PHY layer.

ZigBee is built upon the IEEE 802.15.4, and it describes a communication layer at level 3 in the OSI model that offers network topology creation, authentication and security aimed for low-power and low-rate applications (ZigBee, 2013) & (Krishnamachari, 2004).

ZigBee defines the network layer, the application framework layer, and the application profile layer as shown in figure 1. The network layer handles routing and the mesh capability. The application profile assists in executing set of actions and a common data exchange protocol. Security aspects can be described in application framework layer, however due to its flexibility it can be applied in any stack layer of ZigBee.

## IV. Factors that Influence the Designing of WSN Systems

There are several factors that influence the design of WSN systems. These factors are power/energy consumption, environment that the system operates on, hardware constraints, transmission media, scalability, topology and fault tolerance. In designing WSN protocols, algorithms and prototypes these factors are much being considered.

*A. Power Consumption*

A WSN node encompasses power that is limited (depending on battery life) and difficulty to renew in some situations. Failure of one or several sensor nodes in a network can have a major impact that might cause change in a WSN topology. Power efficiency in WSN systems has a huge effect on the lifetime of that WSN system. Moreover, power consumption also has a relationship with the type of sensing application. The application that requires constant monitoring/sensing might consume much power that the one that require infrequent monitoring.

*B. Fault Tolerance*

Wireless sensor nodes might be prone to failure due to harsh environment, shortage of power or physical destruction. Appropriate WSN designing should consider reliability such that, failure of one or more sensor node(s) should not affect the whole WSN. The level of fault tolerance differs according to application designed and environment. The reliability of a sensor node is given by the following mathematical model developed by (Hoblos, 2000).

$$R_p\ (t) = exp\ (-\lambda_p\ t)$$

Where:

$\lambda_p$ - Failure rate of sensor node $p$

$t$ - Time period

$R_p\ (t)$ – Reliability of node $p$ at time $t$

*C. Topology*

Topology defines the way in which the sensor nodes are arranged. There are several ways of deploying sensor nodes in the field. That includes; placing sensor nodes one by one using hand (or robot), dropping sensor nodes from the plane and throwing by a catapult.

The way sensor nodes are arranged in the field after deployment determines the network and communication performance along with the lifetime of the WSN. The topology might change when sensor nodes changes, when nodes fails due to lack of power or when the sensor nodes' on/off schedule is changed.

*D. Scalability*

The number of sensor nodes employed in the field can range from tens to hundreds to millions varying on the application type and environment. However, this brings another challenge of having efficient protocols and algorithms that can handle such large number of sensor nodes while being energy-efficient to maximize the network lifetime of the WSN.

*E. Environment that operate on*

Sensor nodes can operate in many diverse environments. These include the following environments as summarized with their respective applications in Table 1.

TABLE 1
DIVERSITY ENVIRONMENT THAT CAN WSN CAN OPERATE ON

| Environment | Application |
|---|---|
| Automobiles | Security, vehicle tracking |
| Farms | Monitoring conditions that affects farming and livestock |
| Forest | Fire detection |
| Home | Home automation |
| Hospitals | Tracking patients, doctors, administration of drugs |
| Human body | Telemedicine applications |
| Industries | Monitoring products |
| Military and Battlefields | Attack detection, surveillance |
| Museums | Localization, security, interactive environment |
| Office buildings | Temperature control |
| Satellite | Spatial complexity of plant species |
| Soil | Precision agriculture |
| Underwater | Monitoring water quality, studying behaviors of aquatic animals |
| Water bodies | Flood detection |

**Table 1**: Diversity environments that WSN can operate on.

*F. Hardware constraints*
There are 4 basic parts that comprises the wireless sensor node namely; power unit, processor, sensor unit and transceiver.

i.   *Processor*. This part manages every module of the sensor mode. A WSN processor may comprise of memory that can be attached on-board or combined on the board.

ii.  *Sensor Unit.* This part comprises of several sensors for data acquisition in the area of application. The data collected from the sensors are according to the parameters used in that particular application like pH, temperature and conductivity. The sensor unit is normally combined with the analog-to-digital converter (ADC) that converts the sensor's node analog signals to digital, which is directed into the processor.

iii.   *Power Unit*. This is the power of the sensor node. Normally battery is used but solar energy is also possible to be used as an alternative to battery. This unit powers each part of the sensor node that requires energy.

iv.   *Transceiver*. This part of a sensor node deals with communication. Communication between any two wireless sensor nodes is done by a transceiver.

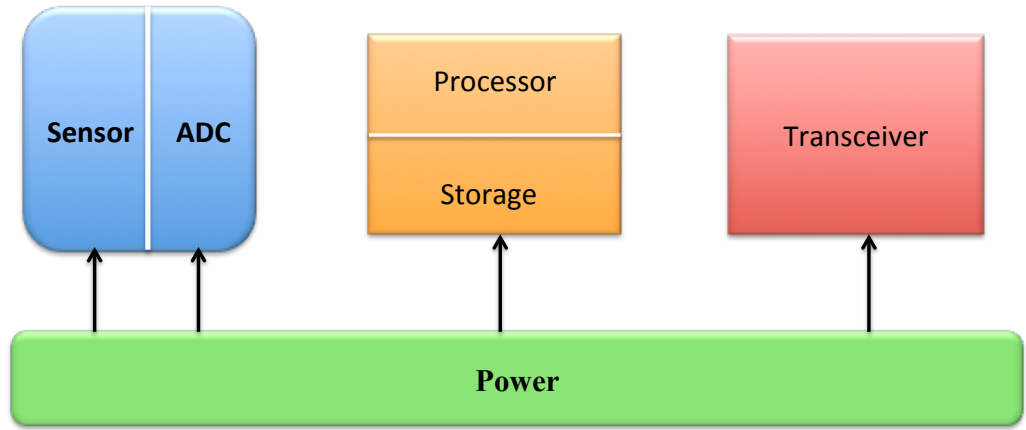

**Figure. 2.** Major components of a sensor node

## V.   Proposed Prototype – IEEE 802.15.4 Based Autonomous Water Quality Monitoring Prototype

The proposed Autonomous Water Quality Monitoring Prototype (AWQMP) is comprised of one or more super node(s) and numerous ordinary nodes. Since power conservation is one of the major challenges in designing WSN systems, solar panels are used in AWQMP. All super and ordinary nodes are attached with solar panels. However, the super node is attached to a large number of sensor nodes and with a more powerful solar panel. Long distance Ethernet radio is used by the super node for data transmission to the base station at the shore while ordinary nodes uses low-power ZigBee radio to communicate and transmit data between themselves.

AWQMP is designed to use Equalized Cluster Head Election Routing Protocol, ECHERP (Nikolidakis, 2013) routing protocol. This is because studies that have compared the performance of WSN routing protocols have shown that ECHERP is better than other protocols like Low Energy Adaptive Clustering Hierarchy (LEACH) (Heinzelman, 2000), Power-Efficient GAthering in Sensor Information Systems (PEGASIS) (Raghavendra, 2002) and Base-Station Controlled Dynamic Clustering Protocol (BCDCP) (Muruganathan, 2005) when considering the number of nodes that remain alive and energy dissipation.

All the nodes are dynamically gathered into clusters. From each clusters, one node is chosen to be the Cluster Head (CH). The CHs that are close to the Base Station (BS) can transmit data directly, while the ones that are far away from BS transmits data to the closest CH. BS is assumed to be fixed and has unlimited power. CHs are chosen by using Gaussian elimination method (Rusell, 2012), where by the nodes that minimizes the most the use of total energy are elected in turns to be CHs as shown in figure 2. Figure 3 describes an example of actual deployment of AWQMP.

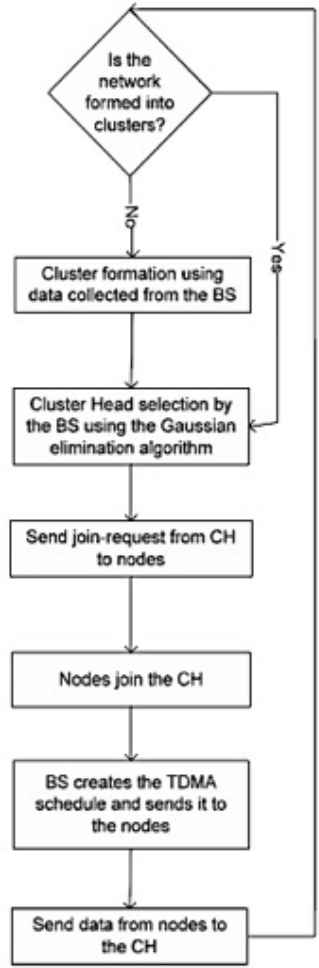

**Figure. 3.** Cluster formation and data sending in ECHERP

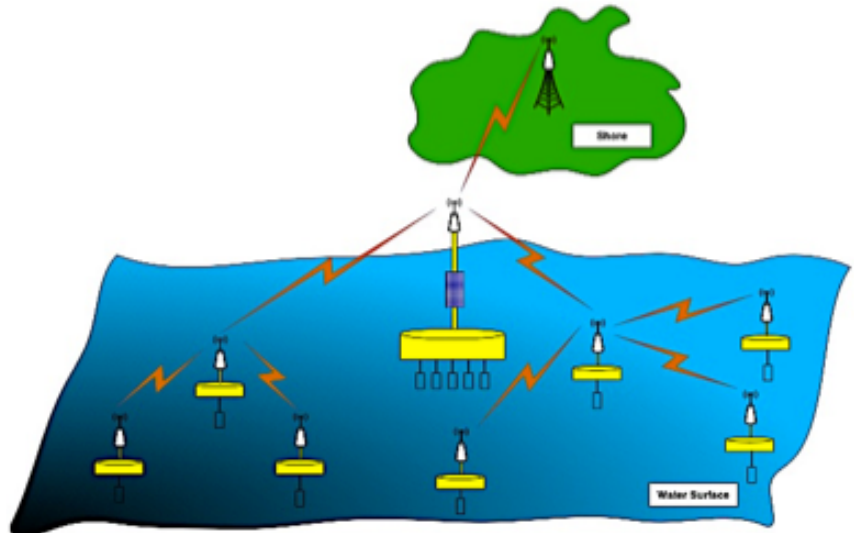

**Figure. 4.** Example of the actual AWQMP deployment

## VI.    Hardware Design

In this section we presents the hardware design, sensors and solar unit to be used in AWQMP.

*A. Wireless Sensor Node*

AWQMP sensor node is comprised of sensor unit and microprocessor unit sharing a solar power via solar panel as a source of energy as shown in the figure 4. AWQMP is designed to use controller from Cirronet (2013) and CC2430 IC from Texas Instrument (2013).

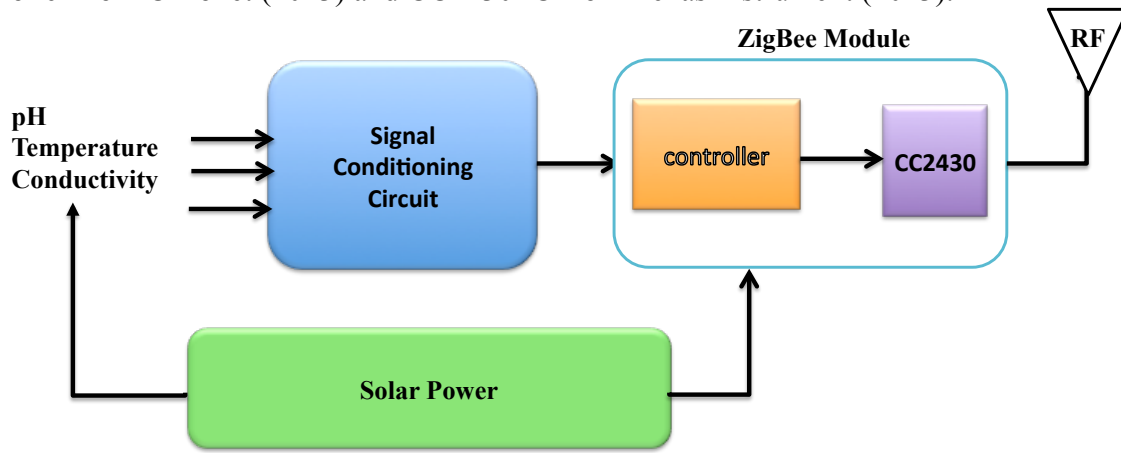

**Figure. 5.** Block diagram of IEEE 182.15.4 based wireless sensor node

*B. Sensor Unit*

A sensor unit in AWQMP is a unit comprises of 3 sensors (pH, temperature and conductivity) that are used to determine the water quality.

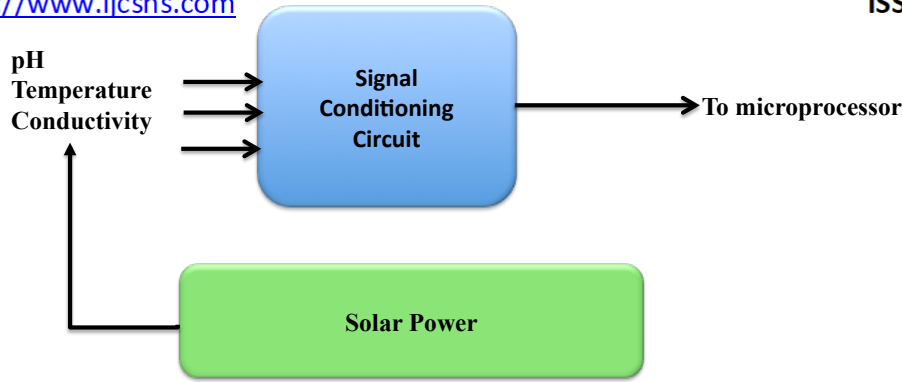

**Figure. 6.** Block diagram of the AWQMP Sensor Unit

   The data acquired by the sensors are converted into electronic signals and transmitted through Signal Conditioning Circuit (SCC) to the microprocessor that convert it to the understandable information shown in figure 5.

i.   *pH*: Water pH is the measure of concentration of H+ ions in the water. As water pH decreases, it becomes more acidic and as the number of H+ ions increases, it becomes more alkaline. The AWQMP is designed to use YSI pH 6-series sensors (YSI, 2013) for pH measurements. The sensors can handle all ionic strength conditions, from seawater, to "average" freshwater lakes and rivers, to pure mountain streams. The sensors specifications are: a range of 0 to 14 units with resolution of 0.01 units and accuracy of ±0.2 unit.

ii.  *Temperature:* The temperature sensor used is YSI 5500D. The sensor temperature range is -15 to 70°C (5 to 158°F) with accuracy of ±0.3°C and resolution of 0.1°C.

iii. *Conductivity*: Electrical conductivity is a good sign of water quality. YSI pro2030 sensors are used. The sensors can measures the conductivity in the range 0-200 ms/cm (auto range) with accuracy of 1-m or 4-m cable, ±1.0% of reading or 1.0 us/cm (whichever greater) and a resolution of 0.0001 to 0.1 ms/cm (range dependent).

| Sensor | Manufacture | Model | Specification |
|---|---|---|---|
| **pH** | YSI | 6-Series | Range: 0-14 units<br>Resolution: 0.01 units<br>Accuracy: ±0.2 units |
| **Temperature** | YSI | 5500D | Range: 15-70°C<br>Resolution: 0.1°C<br>Accuracy: ±0.3°C |
| **Conductivity** | YSI | Pro2030 | Range:0-200mS/cm<br>Resolution: 0.1 mS/cm<br>Accuracy: ±1.0% |

**Table 1**. Sensor specifications for AWQMP

*C. Solar Unit*

In underwater environment, designing the power source for the IEEE 802.15.4 based system can be done through physical wires from the shore, battery or solar energy. However, using wires to connect the sensor nodes is not sensible due to the way sensor nodes are dispersed and expensiveness in implementation.

The system can be designed to use batteries for its sensor nodes, however its great drawback it is the lifetime. The energy in batteries will exhaust and the sensor nodes will die after a certain time. Moreover, replacing exhausted sensor nodes batteries repeatedly is a tiresome and difficulty task.

To make the system last long while avoiding unnecessary tasks, AWQMP is designed to use solar energy. Solar panels are used to supply the energy to the sensor nodes. Also accumulator is attached with the system for recharging in the times when solar energy is not sufficient.

*D. Base Unit*

AWQMP's base station is designed to use IEEE 182.15.4 module automated as a controller that collects data from sensor nodes in water. The base station in the shore is designed to have unlimited power. Data collected from the sensor nodes is channeled to the PC by RS 232 protocol and showed in the BS Graphical User Interface (GUI).

VII.      **Software Design**

AWQMP is designed to use Arduino Mega 2560, an open-source electronic prototyping platform that allows to create interactive electronic objects programmed by Arduino IDE (Arduino, 2013).

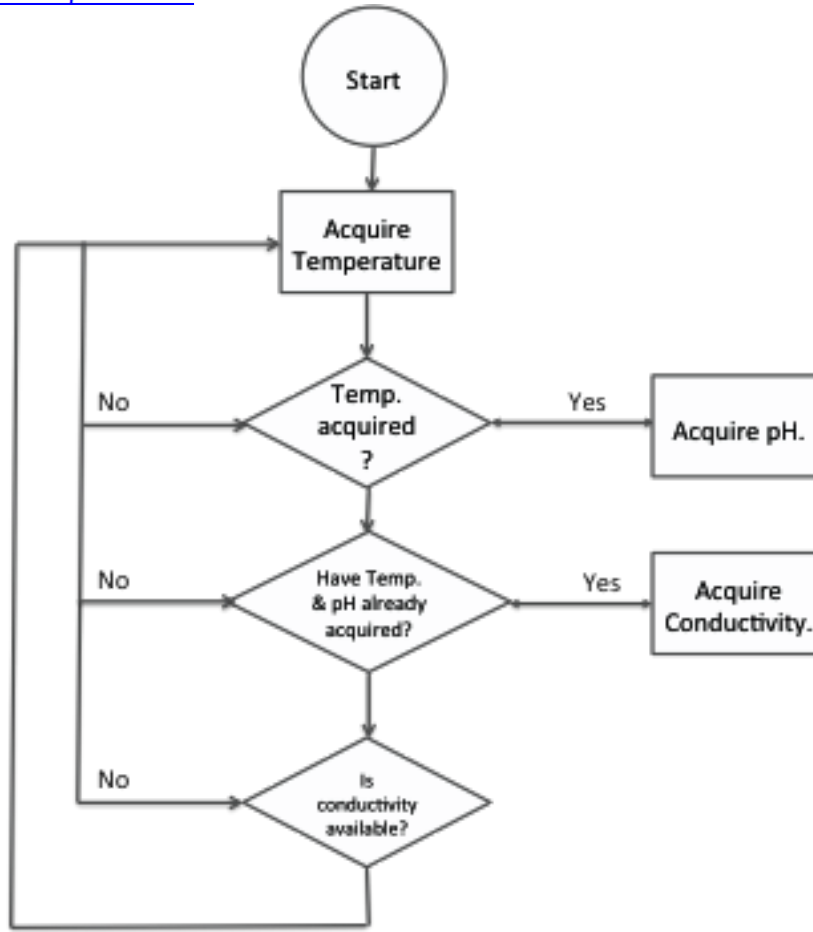

**Figure 7.** Data acquisition process by Arduino from Temperature, pH and Conductivity sensors.

Data acquisition follows the process shown in figure 6. Arduino boots up, serial port initializes with pH, conductivity and temperature sensor circuits to connect with Arduino. Arduino waits for predefined amount of time (usually 5 seconds) and starts collecting data from the sensors. First, Arduino acquire temperature data and then asks for pH data. If temperature data is not acquired in a predefined amount of time, Arduino quit that cycle and start again. Next upon acquiring pH data, it acquires conductivity data and sends the data to the computer. Also, if pH data is not acquired in a predefined amount of time, Arduino quit that cycle and start again.

## VIII.    Conclusion

In this paper, the design of a novel prototype based on IEEE 802.15.2.4 and solar energy for water quality monitoring is described. The prototype used ECHERP routing protocol for energy conservation purposes and solar panels are used instead of batteries to ensure the system will last in a long period of time.  The proposed prototype provides a cost-effective approach in good water quality monitoring, environmental protection and tracking pollution sources compared to manual approaches. Moreover, from data analysis acquired from this prototype will assist in understanding the behavior of aquatic animals in implemented water bodies.